\documentclass[preprint,showpacs,preprintnumbers,amsmath,amssymb,aps,nofootinbib]{revtex4}

\usepackage{graphicx}% Include figure files
\usepackage{dcolumn}% Align table columns on decimal point
\usepackage{bm}% bold math

\begin{document}

\title{Polarized Deep Inelastic and Elastic Scattering From Gauge/String Duality}

\author{Jian-Hua Gao}
\email{gaojh79@ustc.edu.cn} \affiliation{Department of Modern
Physics, University of Science and Technology of China, Hefei,
Anhui 230026, People's Republic of China} \affiliation{Department
of Physics, Shandong University, Jinan, Shandong, 250100, People's
Republic of China}

\author{Bo-Wen Xiao}
\email{bxiao@lbl.gov}
\affiliation{Nuclear Science Division, Lawrence Berkeley National Laboratory, Berkeley, CA 94720, USA}
\date{\today}

\begin{abstract}
In this paper, we investigate deep inelastic and elastic
scattering on a polarized spin-$\frac{1}{2}$ hadron using
gauge/string duality. This spin-$\frac{1}{2}$ hadron corresponds
to a supergravity mode of the dilatino. The polarized deep
inelastic structure functions are computed in supergravity
approximation at large t' Hooft coupling $\lambda$ and finite $x$
with $\lambda^{-1/2}\ll x<1$. Furthermore, we discuss the moments
of all structure functions, and propose an interesting sum rule
$\int_{0}^{1} \textrm{d}x g_2\left(x, q^2\right) =0$ for $g_2$
structure function which is known as the Burkhardt-Cottingham sum
rule in QCD.  In the end, the elastic scattering is studied and
elastic form factors of the spin-$\frac{1}{2}$ hadron are
calculated within the same framework.
\end{abstract}

\maketitle

\section{Introduction}
Gauge/string
duality\cite{Maldacena:1997re,Witten:1998qj,Gubser:1998bc}
provides us new insights into gauge theories in strong coupling
regime. According to the gauge/string duality, the dual string
theory, which corresponds to conformal gauge theories(e.g., the
$\mathcal{N}=4$ super Yang Millis theory), is embedded in $AdS_5
\times S^5$ space with the metric
\begin{equation}
ds^{2}= g_{M,N} \textrm{d}X^{M}\textrm{d}X^{N} =\left(\frac{r^{2}}{R^2}\eta_{\mu \nu} \textrm{d}y^{\mu}\textrm{d}y^{\nu}+\frac{R^2}{r^2}\textrm{d}r^{2}\right)+R^2\textrm{d}\Omega_5^2,
\end{equation}
where $g_{M,N}$ is the ten-dimensional metric and $\eta_{\mu
\nu}=\left(-,+,+,+\right)$ is the mostly plus flat space metric.
Here we use $M, N$ as indices in ten dimensions, $m,n$ as indices
in $AdS_5$ and $\mu,$ $\nu$ as those in four dimensional flat
space which lives on the boundary of the $AdS_5$ space. $R$, which
is the curvature radius of the $AdS_5$ space, is also equal to the
radius of the five-sphere $S^5$. It is given by the duality
\begin{equation}
R^2 = l_{s}^2\sqrt{4\pi g_{st}N}
\end{equation}
where the string coupling $g_{st}$ and the string length $l_s$ are
given by $4\pi g_{st} =g_{YM}^2$ and $l_s^2=\alpha^{\prime}$ with
$\alpha^{\prime} $ being the regge slope parameter, respectively.
The t'Hooft coupling is defined as $\lambda=g_{YM}^2 N= 4\pi
g_{st} N$. One can easily see that the large t'Hooft coupling
limit is equivalent to the limit $R^2 \gg l_s^2$. In the limit
$g_{st}\ll 1$ and $R^2 \gg l_s^2$, the string theory can be
approximated by supergravity. Then the duality reduces to
correspondence between gauge theories at large t'Hooft coupling
and supergravity, One can investigate the nonperturbative
properties of gauge theories at large t'Hooft coupling by studying
the corresponding supergravity theory. There are also some
interesting connections between the Type $\textrm{II-B}$
superstring theory and the $\mathcal{N}=4$ super Yang Millis (SYM)
theory. First, the $SU(4)$ $\mathcal{R}$ symmetry of the
$\mathcal{N}=4$ SYM is the $SO(6)$ isometry of $S^5$. Furthermore,
the $SO(4,2)$ conformal symmetry of the gauge theory is the
isometry of $AdS_5$. In addition, there is an implication that the
radial direction ($r$) in $AdS_5$ can be identified with the
energy scale in four dimensional SYM theory, namely, $E\sim
\frac{r}{R^2}$.

There have been substantial progresses in studying strong coupling
gauge theories especially in terms of deep inelastic scattering. A
few years ago, Polchinski and Strassler\cite{Polchinski:2001tt,
Polchinski:2002jw} studied the deep inelastic scattering on
hadrons by using gauge/string duality where the usual structure
functions $F_1$ and $F_2$ are calculated for both spinless and
spin-$\frac{1}{2}$ hadrons when Bjorken-$x$ is finite
($\lambda^{-1/2}\ll x<1$) where supergravity approximation is
valid. The spinless hadron and spin-$\frac{1}{2}$ hadron
correspond to supergravity modes of dilaton and dilatino,
respectively. Furthermore, they also investigated the case at
small-$x$ where the Pomeron contribution with a trajectory of
$2-\mathcal{O}\left(\frac{1}{\sqrt{\lambda}}\right)$ was found.
Since an infrared cutoff $\Lambda$ is introduced in order to
generate confinement, the model is then called hard wall model.
There are also some earlier studies\cite{Brower:2002er,
BoschiFilho:2002zs} on high energy scattering in gaug/string
duality. There have been a lot of further developments along this
direction\cite{Brower:2007xg,Brower:2006ea,BallonBayona:2008zi,BallonBayona:2007rs,BallonBayona:2007qr,Pire:2008zf,Cornalba:2008sp}.
A saturation picture based on deep inelastic scattering in AdS/CFT
is developed\cite{Hatta:2007he} afterwards and recently reviewed
in Ref.~\cite{Hatta:2008zz}.  In addition, the deep inelastic
scattering off the finite temperature plasma in gauge/string
duality is recently studied in Refs.~\cite{Hatta:2007cs,
Mueller:2008bt}.

Our main object in this paper is to extend the calculation of deep inelastic scattering on
a spin-$\frac{1}{2}$ fermion in hard wall model,
and compute the parity violating structure function $F_3$
as well as the polarized structure functions $g_1$, $g_2$, $g_3$, $g_4$ and $g_5$.
Among these five polarized structure functions, $g_3$, $g_4$ and $g_5$ are parity violating structure functions.

Type $\textrm{II-B}$ superstring theory, which lives in a ten-dimensional space (e.g., $AdS_5 \times S^5$),
is a parity-violating theory. It contains massless left-handed Majorana-Weyl gravitinos and massless
right-handed Majorana-Weyl dilationos. The gravitino, which is a spin-$\frac{3}{2}$ fermion,
is the superpartner of the graviton. Likewise, the dilatino, which is a spin-$\frac{1}{2}$ fermion,
is the superpartner of the dilaton. It has been proved that type $\textrm{II-B}$ superstring theory
is anomaly free\cite{AlvarezGaume:1983ig, Schwarz:2001sh} in terms of local (gauge) symmetries.
Here we expect that the currents in the dual gauge theory is conserved at finite-$x$ as we will show later in the paper.
In this paper, we focus on the spin-$\frac{1}{2}$ dilatino and calculate its structure functions as well as form factors.
In order to study the polarized structure functions and form factors, we follow the set-up in Ref.~\cite{Polchinski:2002jw}
and assume the dilatino has a small mass $M$ which eventually can be related to the cutoff scale $\Lambda$.

This paper is organized as follows. In section \ref{deft}, we
provides the definitions for various structure functions as well
as kinematic variables. In section \ref{hard}, we calculate the
expectation value of the $\mathcal{R}$-currents in our gedanken
experiment of polarized deep inelastic scattering from
gauge/string duality. This eventually leads to the structure
functions at finite $x$. The section \ref{disc} is devoted to the
discussions and comments on the structure functions and their sum
rules. In section \ref{elas}, we focus on the elastic scattering
and derive the form factors for the spin-$\frac{1}{2}$ hadron.
Finally in section \ref{conc}, we summarize our results.

\section{Polarized Deep inelastic Scattering}
\label{deft}
\begin{figure}[htbp]
\includegraphics[width=10cm]{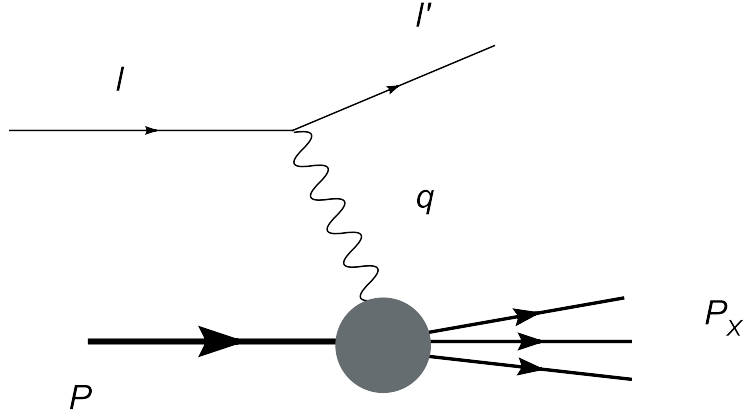}
\caption{Illustration of DIS.}
\label{DIS}
\end{figure}

The hadronic tensor $W^{\mu\nu}$ is defined as
\begin{equation}
  W^{\mu\nu} = \int \! d^4 \xi \, e^{i q{\cdot}\xi} \,
  \langle P, \mathcal{Q}, S |[ J^\mu(\xi), J^\nu(0)] | P, \mathcal{Q}, S \rangle \, ,
  \label{dis3}
\end{equation}
with $J^{\mu}$ being the incident current. The hadronic tensor $W_{\mu\nu}$ can be split as
\begin{equation}
  W_{\mu\nu} =
  W_{\mu\nu}^\mathrm{(S)}(q, P) + i \, W_{\mu\nu}^\mathrm{(A)}(q;P,S) \, ,
  \label{dis10}
\end{equation}
According to Lorentz and CP invariance, the symmetrical and
antisymmetrical parts can be expressed in terms of 8 independent
structure functions as\cite{Anselmino:1994gn,
Lampe:1998eu},\footnote{There are some sign changes in our
definition comparing to the usual definition in
\cite{Anselmino:1994gn, Lampe:1998eu}. These sign changes arise
due to reason that we use the most plus metric throughout this
paper instead of the usual most minus metric.}
\begin{eqnarray}
 W_{\mu\nu}^\mathrm{(S)} &=&
 \left( \eta_{\mu\nu}-\frac{q_\mu q_\nu}{q^2}\right) \left[F_1(x,q^2)+\frac{M S\cdot q}{2P\cdot q}g_5(x,q^2)\right]\nonumber\\
& &  - \frac{1}{P{\cdot}q}\left( P_\mu - \frac{P{\cdot}q}{q^2} \, q_\mu \right)\left( P_\nu - \frac{P{\cdot}q}{q^2} \, q_\nu \right)
  \left[F_2(x,q^2)+\frac{M S\cdot q}{P\cdot q}g_4(x,q^2)\right]\nonumber\\
& &  -\frac{M}{2P\cdot q}\left[\left( P_\mu - \frac{P{\cdot}q }{q^2}q_\mu\right)\left(S_\nu - \frac{S{\cdot}q}{P{\cdot}q} \, P_\nu\right)
+\left( P_\nu - \frac{P{\cdot}q }{q^2}q_\nu\right)\left(S_\mu - \frac{S{\cdot}q}{P{\cdot}q} \, P_\mu\right)\right]g_3(x,q^2)\nonumber\\
  W_{\mu\nu}^\mathrm{(A)}
  &=&
  -\frac{M \, \varepsilon_{\mu\nu\rho\sigma} \, q^\rho}{P{\cdot}q}
  \left\{
    S^\sigma \, g_1(x,q^2) +
    \left[ S^\sigma - \frac{S{\cdot}q}{P{\cdot}q} \, P^\sigma \right] g_2(x,q^2)
  \right\} -\frac{\varepsilon_{\mu\nu\rho\sigma}q^\rho P^\sigma}{2P{\cdot}q}
  F_3(x,q^2),
\label{wmunu}
\end{eqnarray}
where $M$ is the mass of the hadron, $S$ is its polarization, $q$
is the momentum carried by the current and $P$ is the initial
momentum of the hadron (See Fig.~(\ref{DIS}).). In deep inelastic
scattering, we define the kinematic variables as the following:
\begin{equation}
x=-\frac{q^2}{2P\cdot q} \quad \textrm{and} \quad P_X^2=(P+q)^2.
\end{equation}
The mass of the intermediate state after the scattering is defined
as $M_X^2=s=-P_X^2$. All the structure functions are functions of
$x$ and $q^2$.
\section{Polarized structure functions in hard wall model}
\label{hard} In the so-called hard wall model, Polchinski and
Strassler impose a confinement scale $\Lambda$ in the fifth
dimension of $AdS_5$ space. As we will see later in the paper (see
eq.~(\ref{mas})), this scale also provides a mass scale for the
hadrons. Following Polchinski and
Strassler\cite{Polchinski:2002jw}, we perform a gedanken
experiment of polarized deep inelastic scattering which occurs
between the boundary and the cutoff scale $\Lambda$. Here we first
summarize their set-up before we extend the calculations to the
polarized case.

The incident current is chosen to be the $\mathcal{R}$-current
which couples to the hadron as an isometry of $S^5$. According to
the AdS/CFT correspondence, the current excites a nonnormalizable
mode of a Kaluza-Klein gauge field at the Minkowski boundary of
the $AdS_5$ space
\begin{equation}
\delta G_{m a} = A_m(y,r) v_a(\Omega) ,
\end{equation}
where $v_a(\Omega)$ denotes a Killing vector on $S^5$ with
$\Omega$ being the angular coordinates on $S^5$. $ A_m(y,r)$ is
the external potential in the gauge theory corresponding to the
operator insertion $n_\mu J^\mu(q)$ on the boundary of the fifth
dimension of the $AdS_5$ space with the boundary condition
\begin{equation}
A_\mu(y,\infty) =  A_\mu(y)|_{\rm 4d}= n_\mu e^{i q \cdot y}.
\end{equation}
This gauge field fluctuation $ A_m(y,r)$ can be viewed as a vector
boson field which couples to the $\mathcal{R}$-current $J^{\mu}$
on the Minkowski boundary, and then propagates into the bulk as a
gravitational wave, and eventually interacts with the supergravity
modes of the dilatino or dilaton. The gauge field satisfies
Maxwell's equation in the bulk, $D_m F^{mn} = 0$. With a gauge
choice, one can solve this equation for $A_\mu$. Usually people
choose the gauge $A_r =0$. However, the problem is easier in the
Lorenz-like gauge, $i\eta^{\mu\nu} q_\mu A_\nu + R^{-4} r
\partial_r (r^3 A_r) = 0$.
With given boundary conditions, one obtains the solution
\footnote{Here we have corrected a minus sign typo in the solution
of $A_r$ in Ref.~\cite{Polchinski:2002jw}.}
\begin{eqnarray}
A_\mu &=& n_\mu e^{iq\cdot y} \frac{qR^2}{r} K_1(qR^2/r)  \ ,\\
A_r &=& - i q \cdot n e^{iq\cdot y} \frac{ R^4}{r^3} K_0(qR^2/r)\ .
\end{eqnarray}
where $q=\sqrt{q^2}$(Note that in $-+++$ metric signature, $Q^2=q^2>0$ for spacelike current.).
Since $K_{n} (qR^2/r) \sim \exp \left(-qR^2/r\right)$, the deep inelastic scattering should be
localized around $r_{\textrm{int}} \simeq q R^2$ which is far away from the cutoff $r_0=\Lambda R^2$
for hard scattering when $q^2 \gg \Lambda^2$.

Spin-$\frac{1}{2}$ hadrons correspond to supergravity modes of the
dilatino. In the conformal region, one can write the dilatino
field as
\begin{equation}
\lambda = \psi(y,r) \otimes \eta(\Omega)\ ,
\end{equation}
where $\psi(y, r)$ is an $SO(4,1)$ spinor on $AdS_5$ and
$\eta(\Omega)$ is a normalized $SO(5)$ spinor on $S^5$. The
wave-function $\psi$ satisfies a five-dimensional Dirac equation
\footnote{We also noticed that there are some typos in the Dirac
equations in Ref.~\cite{Polchinski:2002jw} where there is an extra
$i$ in Eq.~(\ref{dirac0}) while an $i$ is missing in
Eq.~(\ref{dirac}). The detailed derivation is provided above.}
\begin{equation}
\label{dirac0}
-D\hspace{-7.7pt}\slash \hspace{1.1pt}\psi = {m} \psi\ .
\end{equation}

The solution to this Dirac equation is\cite{Mueck:1998iz},
\begin{equation}
\psi = e^{i p \cdot y} \frac{C'}{r^{5/2}} \Bigl[ J_{mR - 1/2}(M
R^2/r) P_+ + J_{mR + 1/2}(M R^2/r) P_- \Bigr] u_\sigma\ ,
\label{dso}
\end{equation}
where
\begin{equation}
p\hspace{-5.3pt}\slash u_\sigma = i M u_\sigma\quad (\sigma = 1,2)\ ,\quad
M^2 = -p^2\ ,\quad P_{\pm} = \frac{1}{2} ( 1 \pm \gamma^{5} )\ . \label{dirac}
\end{equation}
Here we define the $\gamma$-matrices according to the Dirac algebra in the mostly plus metric signature $-+++$ (see, e.g., the notation in Ref.~\cite{Weinberg:1995mt})
\begin{equation}
\{\gamma^{\mu},\gamma^{\nu}\}\equiv \gamma^{\mu}\gamma^{\nu}+\gamma^{\nu}\gamma^{\mu}=2\eta^{\mu \nu} \times \mathbf{1}_{4\times 4},
\end{equation}
which gives an additional factor of $-i$ in $\gamma_{\mu}$
($\mu=0,1,2,3$). It is straightforward to see that this explains
the factor $i$ in front of the fermion mass in Eq.~(\ref{dirac})
and the $\gamma^5$ is the same as the usual definition.

The dilatino is taken to be in a charge eigenstate with charge
$\mathcal{Q}$ under the $U(1)$ symmetry, which yields
$v_{a}\partial ^{a} \eta(\Omega) =i\mathcal{Q} \eta(\Omega)$. This
$U(1)$ symmetry arises from the $U(1)$ subgroup of the $SU(4)$
$\mathcal{R}$-symmetry.

For the initial hadron, by assuming $MR^2 /r \ll 1$ in the interaction region and expanding the Bessel functions in Eq.~(\ref{dso}) up to linear term in $M$, one gets
\begin{equation}
\psi_{\rm i} \approx e^{i P \cdot y} \frac{c'_{\rm
i}}{\Lambda^{3/2} R^{9/2} } (\frac{r_0}{r})^{ mR + 2} \left[P_+
u_{\rm i\sigma}+\frac{M r_0}{2(mR+1/2)\Lambda r} P_- u_{\rm
i\sigma}\right].
\end{equation}
\begin{equation}
\bar{\psi}_{\rm i} \approx e^{-i P \cdot y} \frac{c'_{\rm i}}{\Lambda^{3/2} R^{9/2} }
(\frac{r_0}{r})^{ mR + 2} \left[ \bar{u}_{\rm i\sigma}P_- +\frac{M r_0}{2(mR+1/2)\Lambda r}  \bar{u}_{\rm i\sigma}P_+\right].
\end{equation}
In order to obtain polarized contribution of structure function, we have kept the next leading order $M$ of initial hadron. The conformal dimension $\Delta$ of the state is found to be $mR+2$.
For the intermediate hadron, $M_X \gg \Lambda$ and
\begin{equation}
\psi_{X} \approx e^{i (P+q) \cdot y} \frac{c'_{X} s ^{1/4} \Lambda^{1/2}
R^{1/2}}{r^{5/2}}
\Bigl[
J_{mR - 1/2}(M_X R^2/r) P_+ + J_{mR + 1/2}(M_X R^2/r) P_- \Bigr]
u_{X\sigma} \ .
\end{equation}
\begin{equation}
\bar{\psi}_{X} \approx e^{-i (P+q) \cdot y} \frac{c'_{X} s ^{1/4} \Lambda^{1/2}
R^{1/2}}{r^{5/2}}
\bar{u}_{X\sigma}\Bigl[
P_- J_{mR - 1/2}(M_X R^2/r)  + P_+ J_{mR + 1/2}(M_X R^2/r)  \Bigr]
 \ .
\end{equation}

Before getting into the detailed calculation, let us look into the center of mass square of the intermediate states in ten dimension. It is easy to see that
\begin{equation}
\tilde{s} =-g^{M, N} P_{X, M} P_{X, N} \leq \frac{R^2}{r_{\textrm{int}}^2} q^2 \left(\frac{1}{x}-1\right) \quad \textrm{with} \quad r_{\textrm{int}}=q R^2.
\end{equation}
Thus we know that $\alpha^{\prime} \tilde{s}=\frac{1}{\sqrt{\lambda}}\left(\frac{1}{x}-1\right) \ll 1$ when $\frac{1}{\sqrt{\lambda}} \ll x <1$.
In this range of $x$, only massless string states are the relevant intermediate states produced during the interaction,
and supergravity calculation should be valid and reliable to obtain the structure functions. When $x$ gets smaller,
 the massive string modes are excited and string scattering amplitude should be taken into account.
 This has been calculated thoroughly for $F_1$ and $F_2$. Unfortunately, we leave this part of the
 calculation for polarized structure functions for future studies.

Therefore, it is straightforward to compute the matrix element and obtain
\begin{eqnarray}
& & n_\mu \langle {P_X,X,\sigma'}|  J^\mu(0) |{P,\cal{Q},\sigma}\rangle\nonumber\\
&=&
i{\cal{Q}} \int d^{6} x_\perp \sqrt{-g}\, A_m \bar{\lambda}_X \gamma^m \lambda_{\rm i}\\
&=&
i{\cal{Q}} \int d^{6} x_\perp \sqrt{-g}\, \left(A_\mu \bar{\lambda}_X {e^\mu}_{\hat{\mu}}\gamma^{\hat{\mu}} \lambda_{\rm i}
+A_r \bar{\lambda}_X {e^r}_{\hat{r}}\gamma^{\hat{r}} \lambda_{\rm i}\right)
\end{eqnarray}
where $n_{\mu}$ is the polarization of the current $J^\mu$, $\hat{\mu}$ and $\hat{r}$ are the tangent space index, and the vielbein ${e^\mu}_{\hat{\mu}}$ and ${e^r}_{\hat{r}}$ are
given by,
\begin{equation}
{e^\mu}_{\hat{\mu}}=\frac{R}{r}{\eta^\mu}_{\hat{\mu}}\quad \textrm{and} \quad \ {e^r}_{\hat{r}}=\frac{r}{R}.
\end{equation}
Here the vielbein is used to make the product Lorentz invariant due to the fact that the gamma matrices are defined in the flat spacetime. It then follows that,
\begin{eqnarray}
& & n_\mu \langle{P_X,X,\sigma'}|  J^\mu(0) |{P,\cal{Q},\sigma}\rangle\nonumber\\
&=&
i{\cal{Q}} \int d r \frac{r^3}{R^3}R^5\frac{c'_{\rm i}}{\Lambda^{3/2} R^{9/2} }\frac{(\Lambda R^2)^{ mR + 2}}{r^{ mR + 2}}\,
\frac{c'_{X} s ^{1/4} \Lambda^{1/2}R^{1/2}}{r^{5/2}}\nonumber\\
& &\times\left(\frac{qR^2}{r}K_1(qR^2/r) J_{mR - 1/2}(M_X R^2/r)\frac{R}{r}\bar{u}_{X\sigma'} n\hspace{-6pt}\slash P_{+}u_{\rm i\sigma}\right.\nonumber\\
& &\ \ \ -iq\cdot n \frac{R^4}{r^3}K_0(qR^2/r) J_{mR + 1/2}(M_X R^2/r)\frac{r}{R}\bar{u}_{X\sigma'} \gamma^5 P_{+}u_{\rm i\sigma}\nonumber\\
& &\ \ \ +\frac{qR^2}{r}K_1(qR^2/r) J_{mR + 1/2}(M_X R^2/r)\frac{R}{r}\frac{M R^2}{(2mR+1)r}\bar{u}_{X\sigma'} n\hspace{-6pt}\slash P_{-}u_{\rm i\sigma}\nonumber\\
& &\left.\ \ \ -iq\cdot n \frac{R^4}{r^3}K_0(qR^2/r) J_{mR - 1/2}(M_X R^2/r)\frac{r}{R}\frac{M R^2}{(2mR+1)r}\bar{u}_{X\sigma'} \gamma^5 P_{-}u_{\rm i\sigma}\right)
\end{eqnarray}
After changing variables to $z=\frac{R^2}{r}$, one finds
\begin{eqnarray}
& &n_\mu \langle{P_X,X,\sigma'}|  J^\mu(0) |{P,\cal{Q},\sigma} \rangle \nonumber \\
&=&
i{\cal{Q}} c'_{\rm i} c'_X s^{1/4}\Lambda^{\tau-1/2}\int_{0}^{1/\Lambda} \textrm{d} z z^{\tau}\nonumber\\
& &\times\left(q K_1(qz) J_{\tau-2}(M_X z)\bar{u}_{X\sigma'} n\hspace{-6pt}\slash P_{+}u_{\rm i\sigma}
-iq\cdot n K_0(qz) J_{\tau-1}(M_X z)\bar{u}_{X\sigma'} \gamma^5 P_{+}u_{\rm i\sigma}\right.\nonumber\\
& & \ \ \ + q z K_1(qz) J_{\tau-1}(M_X z)\frac{M\bar{u}_{X\sigma'} n\hspace{-6pt}\slash P_{-}u_{\rm i\sigma}}{2(\tau-1)}\nonumber\\
& &\left. \ \ \ -i(q\cdot n) z K_0(qz) J_{\tau-2}(M_X z)\frac{M\bar{u}_{X\sigma'} \gamma^5 P_{-}u_{\rm i\sigma}}{2(\tau-1)} \right)
\end{eqnarray}
where
\begin{equation}
r_0=\Lambda R^2 \quad \textrm{and} \quad \tau=\Delta -1/2=mR+\frac{3}{2}
\end{equation}
Using the following integral results
\begin{eqnarray}
\int_{0}^{\infty} d z z^{\tau} K_1(qz) J_{\tau-2}(M_X z)&=& \frac{2^{\tau -1} M_X^{\tau -2} q }{(M_X^2+q^2)^{\tau}} \Gamma(\tau)                \\
\int_{0}^{\infty} d z z^{\tau} K_0(qz) J_{\tau-1}(M_X z)&=& \frac{2^{\tau -1} M_X^{\tau -1}}{(M_X^2+q^2)^{\tau}} \Gamma(\tau)                     \\
\int_{0}^{\infty} d z z^{\tau+1} K_1(qz) J_{\tau-1}(M_X z)&=& \frac{2^{\tau } M_X^{\tau -1} q}{(M_X^2+q^2)^{\tau+1}}\Gamma(\tau+1)                  \\
\int_{0}^{\infty} d z z^{\tau+1} K_0(qz) J_{\tau-2}(M_X z)&=& \frac{2^{\tau } M_X^{\tau -2}}{(M_X^2+q^2)^{\tau+1}}\left[q^2\Gamma(\tau+1)-(M_X^2+q^2)\Gamma(\tau)\right],
\end{eqnarray}
where the upper limits are approximately set to be $\infty$, we have,
\begin{eqnarray}
&& <{P_X,X,\sigma'}|  J^\mu(0) |{P,\cal{Q},\sigma}> \nonumber\\
&=&
i{\cal{Q}} c'_{\rm i} c'_X s^{1/4}\Lambda^{\tau-1/2}2^{\tau -1} M_X^{\tau -2} (M_X^2+q^2)^{-\tau}\Gamma(\tau)\nonumber\\
& &\times\left(q^2\bar{u}_{X\sigma'} \gamma^\mu P_{+}u_{\rm
i\sigma}-iM_X q^\mu \bar{u}_{X\sigma'}  P_{+}u_{\rm i\sigma}
+\frac{\tau}{\tau-1} \frac{M M_X}{M_X^2+q^2} q^2\bar{u}_{X\sigma'} \gamma^\mu P_{-}u_{\rm i\sigma}\right.\nonumber\\
& &\left.\ \ \ +i \frac{\tau}{\tau-1} \frac{M q^\mu q^2}{M_X^2+q^2} \bar{u}_{X\sigma'}  P_{-}u_{\rm i\sigma}-i \frac{M q^\mu}{\tau-1}\bar{u}_{X\sigma'}  P_{-}u_{\rm i\sigma}\right)
\end{eqnarray}
or its complex conjugate\footnote{Note that terms like $iM_X q^\mu
\bar{u}_{\rm i\sigma} P_{-}u_{X\sigma'}$ do not change sign due to
the fact that $\gamma^0$ is imaginary in the notation that we are
working with.},
\begin{eqnarray}
&& \langle {P,\cal{Q},\sigma} | J^\mu(0) |{P_X,X,\sigma'}\rangle \nonumber\\
&=&
-i{\cal{Q}} c'_{\rm i} c'_X s^{1/4}\Lambda^{\tau-1/2}2^{\tau -1} M_X^{\tau -2} (M_X^2+q^2)^{-\tau}\Gamma(\tau)\nonumber\\
& &\times\left(q^2\bar{u}_{\rm i\sigma} \gamma^\mu P_{+}u_{X\sigma'}-iM_X q^\mu \bar{u}_{\rm i\sigma} P_{-}u_{X\sigma'}
+\frac{\tau}{\tau-1} \frac{M M_X q^2}{M_X^2+q^2}\bar{u}_{\rm i\sigma}\gamma^\mu P_{-}u_{X\sigma'} \right.\nonumber\\
& &\left.\ \ \ + i \frac{\tau}{\tau-1} \frac{M q^\mu q^2}{M_X^2+q^2} \bar{u}_{\rm i\sigma}  P_{+}u_{X\sigma'}-i \frac{M q^\mu}{\tau -1}\bar{u}_{\rm i\sigma}  P_{+}u_{X\sigma'}\right).
\end{eqnarray}
With the help of Eq.~(\ref{dirac}), it is easy to see that
$q_\mu<{P_X,X,\sigma'}|  J^\mu(0) |{P,\cal{Q},\sigma}>=0$ and
$q_{\nu}<{P,\cal{Q},\sigma} | J^\nu(0) |{P_X,X,\sigma'}> =0$ as a
result of current conservation. In fact, with the present NLO
approximation, we can only show that $q_\mu<{P_X,X,\sigma'}|
J^\mu(0) |{P,\cal{Q},\sigma}> \sim M^2/q^2$. Nevertheless, we can
expand the initial wavefunction up to NNLO ($M^2$ order), and find
that $M^2/q^2$ contributions are cancelled by NNLO terms in the
initial wavefunction. If one continues to do this to higher
orders, one can show that the current conservation is true for all
orders of $M^2/q^2$. Moreover, using the recursion relations of
Bessel functions, integrating the $dz$ integral by parts and
requiring the $M/\Lambda$ and $M_X/\Lambda$ to be the zeros of
Bessel functions as we use in the later elastic calculation, one
can show that $q_{\nu}<{P,\cal{Q},\sigma} | J^\nu(0)
|{P_X,X,\sigma'}> =0$ vanishes exactly.

Following Polchinski and Strassler, we also define $T^{\mu \nu} $ as
\begin{equation}
T^{\mu \nu} =i\langle P, \mathcal{Q}, S |T\left(J^\mu(q)J^\nu(0)\right) | P, \mathcal{Q}, S \rangle .
\end{equation}
Its imaginary part can be written as
\begin{equation}
\textrm{Im}T^{\mu \nu} =2\pi^2\sum_{X} \delta \left(M_X^2+(p+q)^2\right) \langle P, \mathcal{Q}, S |J^\nu(0) | P+q, X \rangle\langle P+q, X |J^\mu(0)| P, \mathcal{Q}, S \rangle .
\end{equation}
In large $q^2$ limit, we approximately write $\sum_{X}\delta \left(M_X^2+(p+q)^2\right)\simeq \frac{1}{2\pi M_X \Lambda}$.

Summing over radial excitations and final state spin, but keeping the initial spin, along with the relation $\frac{1}{2\pi } W^{S,A}_{\mu \nu} =2\mathrm{Im} T^{S,A}_{\mu \nu}$ derived from the optical theorem,  yields,
\begin{eqnarray}
{W}_{\mu \nu }^{(S)} &=& \pi A^{\prime} {\cal{Q}}^2(\Lambda^2/q^2)^{\tau-1}x^{\tau+1}(1-x)^{\tau-2}\nonumber\\
& &\times
\left\{\left(\eta_{\mu\nu}-\frac{q_\mu q_\nu}{q^2}\right)\left(\frac{1}{2}+\frac{q\cdot S}{2P\cdot q}M\right)
-\frac{1}{P{\cdot}q}\left( P_\mu - \frac{P{\cdot}q}{q^2} \, q_\mu \right)\left( P_\nu - \frac{P{\cdot}q}{q^2} \, q_\nu \right)\right.\nonumber\\
&&\left.-\frac{M}{2P\cdot q}\left[(P_\mu - \frac{P{\cdot}q}{q^2} \, q_\mu)(S_\nu-\frac{S\cdot q}{q^2} q_\nu)
+(P_\nu - \frac{P{\cdot}q}{q^2} \, q_\nu)(S_\mu-\frac{S\cdot q}{q^2} q_\mu)\right]\right\}\\
&=&\pi A_{0}^{\prime }{\mathcal{Q}}^{2}(\Lambda^{2}/q^{2})^{\tau -1}x^{\tau +1}(1-x)^{\tau -2}  \nonumber \\
&&\times \left\{ \left( \eta _{\mu \nu }-\frac{q_{\mu }q_{\nu }}{q^{2}}%
\right) \left( \frac{1}{2}+\frac{q\cdot S}{2P\cdot q}M\right) \right.
\nonumber \\
&&\left. -\frac{1}{P{\cdot }q}\left( P_{\mu }-\frac{P{\cdot }q}{q^{2}}%
\,q_{\mu }\right) \left( P_{\nu }-\frac{P{\cdot }q}{q^{2}}\,q_{\nu }\right)
\left( 1+\frac{q\cdot S}{P\cdot q}M\right) \right.   \nonumber \\
&&\left. -\frac{M}{2P\cdot q}\left[ (P_{\mu }-\frac{P{\cdot }q}{q^{2}}%
\,q_{\mu })(S_{\nu }-\frac{S\cdot q}{q^{2}}q_{\nu })+(P_{\nu }-\frac{P{\cdot
}q}{q^{2}}\,q_{\nu })(S_{\mu }-\frac{S\cdot q}{q^{2}}q_{\mu })\right]
\right\}
\end{eqnarray}
and
\begin{eqnarray}
{W}_{\mu \nu }^{(A)} &=&\pi A_{0}^{\prime }{\mathcal{Q}}^{2}(\Lambda
^{2}/q^{2})^{\tau -1}x^{\tau +1}(1-x)^{\tau -2}  \nonumber \\
&&\times \left\{ -\frac{\epsilon _{\mu \nu \alpha \beta }q^{\alpha }P^{\beta
}}{2P\cdot q}-\frac{M\epsilon _{\mu \nu \alpha \beta }q^{\alpha }S^{\beta }}{%
2P\cdot q}-\frac{M\epsilon _{\mu \nu \alpha \beta }q^{\alpha }}{2P\cdot q}%
\left( S^{\beta }-\frac{q\cdot S}{P\cdot q}P^{\beta }\right) \left( \frac{1}{2x}\frac{\tau +1}{\tau -1}-\frac{%
\tau }{\tau -1}\right) \right\}\nonumber \\
\end{eqnarray}
where $A^{\prime }=\pi |c_{\mathrm{i}}^{\prime }|^2 |c_{X}^{\prime} |^2 2^{2\tau}\Gamma ^{2}(\tau )$.
To obtain ${W}_{\mu\nu}^{(A)}$, we have used the identity,
\begin{eqnarray}
\epsilon^{\mu\nu\alpha\beta}q_\alpha\left[(q\cdot S)P_\beta-(P\cdot q)S_\beta \right]= q^\mu\epsilon^{\nu\alpha\beta\gamma}P_\alpha q_\beta S_\gamma
    -q^\nu\epsilon^{\mu\alpha\beta\gamma}P_\alpha q_\beta S_\gamma
    -q^2\epsilon^{\mu\nu\alpha\beta}P_\alpha S_\beta
\end{eqnarray}

\begin{figure}[tbp]
\begin{center}
\includegraphics[width=8cm]{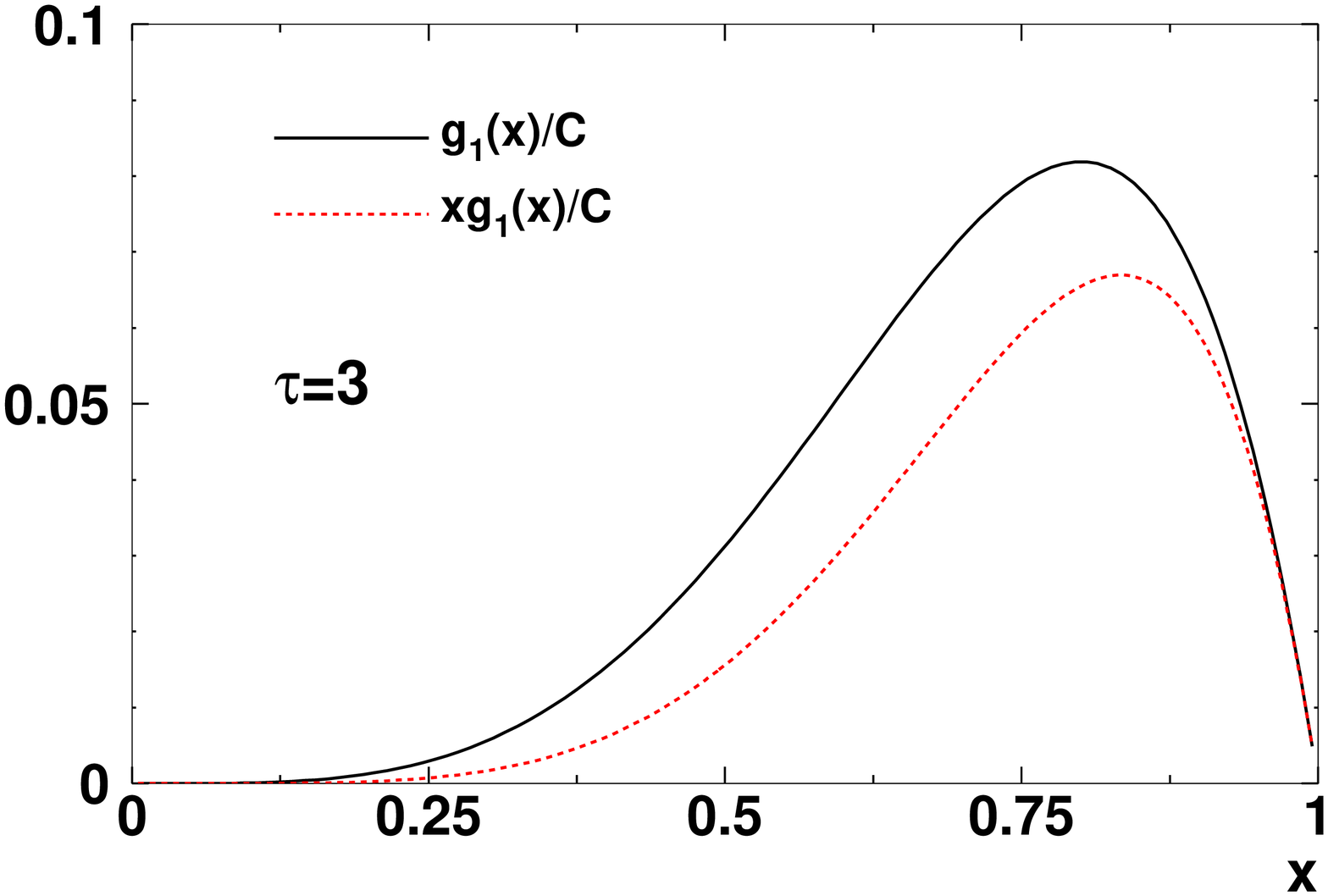}
\includegraphics[width=8cm]{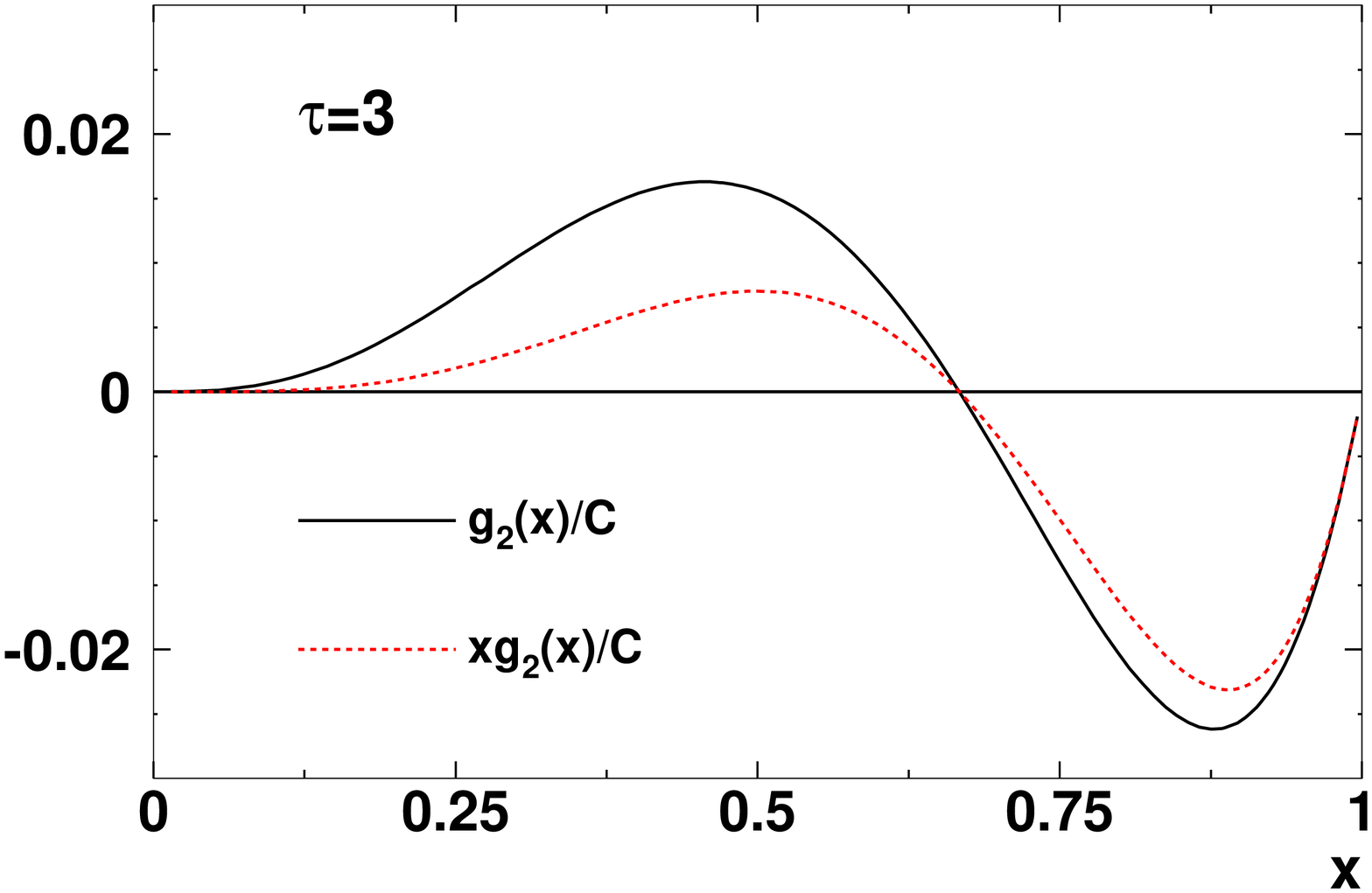}
\end{center}
\caption[*]{Illustration of the $g_1$ and $g_2$ structure functions, where $C=\frac{1}{2}\pi A^{\prime } {\cal{Q}}^2(\Lambda^2/q^2)^{\tau-1}$ and $\tau=3$.}
\label{structure}
\end{figure}

Compare with Eq.~(\ref{wmunu}), we arrive at the final results,
\begin{eqnarray}
 2F_{1} &=&F_{2}=F_{3}=2g_{1}=g_{3}=g_{4}=g_{5}=\pi A^{\prime } {\cal{Q}}^2(\Lambda^2/q^2)^{\tau-1}x^{\tau+1}(1-x)^{\tau-2}\\
 2g_2&=&\left(\frac{1}{2x}\frac{\tau +1}{\tau-1}-\frac{\tau
}{\tau-1}\right)\pi A^{\prime }
{\cal{Q}}^2(\Lambda^2/q^2)^{\tau-1}x^{\tau+1}(1-x)^{\tau-2}.
\end{eqnarray}
The $F_1$ and $F_2$ are exactly the same as the results found in Ref.~\cite{Polchinski:2002jw} by Polchinski and Strassler. The results for $F_3$ and all of the polarized structure functions are new. These structure functions are essentially calculated from the so-called double trace operators with their twist $\tau_p \geq 2$. In figure.~(\ref{structure}), we illustrate the $x$ dependence of the $g_1$ and $g_2$ structure functions. The $F_3$, $g_3$, $g_4$ and $g_5$ structure functions are just twice of the $g_1$. $g_2$ structure functions are especially interesting, it is negative at large $x$ region and positive at relatively small $x$ region which shares the same feature as seen in the proton $g_2$ experiment data.

\section{Discussions}\label{disc}
In this section, we focus on the interpretation of the structure
functions that we obtained from last section using gauge/string
duality. We also compare the results with the structure functions
obtained in QCD for nucleons (For a review in QCD, see e.g.,
Refs.~\cite{Manohar:1992tz,Anselmino:1994gn,Lampe:1998eu,Filippone:2001ux}.).
\begin{itemize}
\item Since only the linear term in $M$ is kept in the initial wavefunction and throughout the calculation, the results shown above are from leading order calculation. The corrections are of order $\frac{M^2}{q^2}$ and $\frac{\Lambda^2}{q^2}$.
\item In QCD, there is an interesting inequality $F_1 \geq g_1$ which is derived from the positivity of the cross section\cite{Manohar:1992tz}. Here we see that $F_1 =g_1$, and the bound is saturated. This indicates that initial hadron is completely polarized. In terms of string theory language, this implies that the struck dilatino just tunnels or shrinks to smaller size of order the inverse momentum transfer during the scattering. As a result, the structure function exhibits a power law behavior in terms of the $q^2$ dependence which comes from the tunneling probability\cite{Polchinski:2001tt, Polchinski:2002jw}.
\item It is now straightforward to compute the moments of all the structure functions when the contributions from $x\ll \lambda^{-1/2}$ are negligible. Typically there are just two different kinds of moments, e.g.,
\begin{eqnarray}
\int_{0}^{1} 2g_1\left(x, q^2\right) x^{n-1}\textrm{d}x&=& \pi A^{\prime } {\cal{Q}}^2(\Lambda^2/q^2)^{\tau-1} \frac{\Gamma\left(\tau-1\right)\Gamma\left(\tau+n+1\right)}{\Gamma\left(2\tau+n\right)} \\
\int_{0}^{1} 2g_2\left(x, q^2\right) x^{n-1}\textrm{d}x&=& \pi A^{\prime } {\cal{Q}}^2(\Lambda^2/q^2)^{\tau-1} \frac{\Gamma\left(\tau-1\right)\Gamma\left(\tau+n\right)}{\Gamma\left(2\tau+n\right)} \frac{1-n}{2}.
\end{eqnarray}
We expect that the moments are correct at least for $n >2$ where the low-x contributions are negligible.
\item In addition, when one sets $n=1$ for $g_2$,
one finds an interesting sum rule
\begin{equation}
\int_0^1 \textrm{d}x g_2\left(x, q^2\right) =0,
\end{equation}
which is completely independent of $\tau$ and $q^2$. In QCD, this sum rule is known as the Burkhardt-Cottingham sum rule\cite{Burkhardt:1970ti} in large $Q^2$ limit. However, this sum rule can be invalidated by non-Regge divergence at low-$x$.
\item Now let us take a closer look at the $n=1$ moment of the $g_1$ structure functions
\begin{equation}
\int_{0}^{1} 2g_1\left(x, q^2\right)\textrm{d}x= \pi A^{\prime } {\cal{Q}}^2(\Lambda^2/q^2)^{\tau-1} \frac{\Gamma\left(\tau-1\right)\Gamma\left(\tau+2\right)}{\Gamma\left(2\tau+1\right)} .
\end{equation}
For sufficiently large $q^2\rightarrow \infty$, this integral vanishes. This contradicts with the naive expectation that $\int_{0}^{1} g_1 \left(x,q^2\right) \textrm{d}x $ should remain finite as $q^2\rightarrow \infty$ since the dilatino has spin-$\frac{1}{2}$.

Before we explain this problem, let us review the case of $F_1$ and $F_2$ structure function\cite{Polchinski:2002jw}. According to energy momentum conservation, the second moment of $F_1$ and the first moment of $F_2$ should have nonzero limit as $q^2\rightarrow \infty$. This is known to be determined by the operator product expansion coefficients of $J^{\mu} J^{\nu}\sim T^{\mu \nu}$. However, it is not true for the result that we found above.
This indicates that some contributions to $F_1$ and $F_2$ which peak around $x=0$ are missing in above calculation. The missing contributions are Pomeron exchanges. At large t'Hooft coupling, the Pomeron exchange is a graviton exchange which yields
\begin{equation}
x F_1 \sim F_2 \propto x^{-1+\mathcal{O}(1/\sqrt{\lambda})}
\end{equation}
at small-$x$ where the correction to the Pomeron intercept arises from the curvature of $AdS_5$.
The Pomeron contribution will survive in the large $q^2$ limit and give us a non-vanishing
second moment of $F_1$\cite{Polchinski:2002jw, Hatta:2007he}.

Therefore, there should be a similar contribution to $g_1$ at
small-$x$. Usually, the physical scattering amplitudes, which can
be written in terms of $F_1+g_1$ and $F_1-g_1$, have the same
leading order $1/x$ singularity. In other words, $g_1$ should
always be less singular than $F_1$. In terms of the Regge theory,
there should be an axial vector Regge exchange contribution which
yields a singular\cite{hatta}
\begin{equation}
g_1\sim \frac{1}{x^{\alpha_{R1}}},
\end{equation}
 with $\alpha_{R1} =1-\mathcal{O}(\frac{1}{\sqrt{\lambda}})$ when $x$ is extremely small.
 This contribution will also survive in large $q^2$ limit and yield a finite first moment.
 This may indicate that most of the hardon spin is carried by the small-$x$ constituents inside the hadron.
\item Normally in QCD, $g_1$ structure function contains two parts, namely, the singlet part
and the non-singlet part. The singlet part contains the polarized singlet quark and gluon spin contributions,
while the non-singlet part can be cast into the Bjorken sum rule. One can subtract off the singlet part
and derive the Bjorken sum rule by calculating $\int_0^1 \textrm{d}x [g_1^{p}\left(x, q^2\right)-g_1^{n}\left(x, q^2\right)]$
at large $Q^2$ limit. Here $g_1^{p}\left(x, q^2\right)$, $g_1^{n}\left(x, q^2\right)$ stand for the $g_1$ structure
functions of proton and neutron, respectively. Since in our above AdS/CFT calculation we only use the $U(1)$
subgroup of $SU(4)$ $\mathcal{R}$-flavor-symmetry group and calculate the contributions from double trace operators,
we can not distinguish the singlet part from the non-singlet part. Both parts are not included in the calculation.
However, if one includes the contribution from the axial currents and uses the full $SU(4)$ group,
then one gets an additional flavor factor $\langle\mathcal{Q}|T^a T^b|\mathcal{Q}\rangle$ where $T^a$ are the $SU(4)$
flavor matrices and the flavor indices $a, b$ are set equal. It is straightforward to see that this flavor factor
also contains both singlet and non-singlet parts. According to our calculation at finite $x$, both of them are small at large $q^2$ limit.
The detail discussions on the small-$x$ limit of the $g_1$ structure function will be available in Ref.~\cite{hatta}.
\footnote{We acknowledge interesting discussions with Y. Hatta, A. Mueller and B. Wu on this issue.}

\item The parity violating structure functions $F_3$, $g_3$, $g_4$ and $g_5$ are as large as the $F_2$
structure function due to the reason that the dilatino is
right-handed fermion in massless limit. They are tightly related
to the peculiar wavefunction of the dilatino. However, we expect
that $g_1$ and $g_2$ may exhibit some common features of the
polarized structure functions of  spin-$\frac{1}{2}$ hadrons in
the non-perturbative region when the coupling is large.
\end{itemize}

\section{Elastic form factors}\label{elas}
In this section, we focus on elastic scattering off a
spin-$\frac{1}{2}$ fermion in gauge/string duality in the hard
wall model framework. In the case of the elastic scattering, the
final state is the same as the initial state which allows us to
set $M_X^2 =M^2$ and $x=1$. Thus, the only variable is $q^2$. In
AdS/QCD model, the meson form factors have been extensively
studied in Refs.~\cite{deTeramond:2008ht, Brodsky:2008pf,
Brodsky:2007hb, Brodsky:2006uqa, Brodsky:2003px,Grigoryan:2007wn,
Grigoryan:2007my,Grigoryan:2007vg,Abidin:2008hn,Abidin:2008ku}.
Furthermore, the nucleon (spin-$\frac{1}{2}$ hadron) form factors
are then computed in Ref.~\cite{Hong:2007dq, Brodsky:2008pg,
Abidin:2009hr}. Here in this section, we would like to follow the
formalism that we developed for the deep inelastic scattering, and
use it in the elastic scattering, then calculate all possible form
factors for spin-$\frac{1}{2}$ hadrons. Here in this section, we
need to keep the full dilatino wavefunction since $q^2/M^2$ is no
longer a large parameter.

To compute the form factors, one can first write down the most general definition for elastic form factors
\begin{eqnarray}
& &\langle {P_X,\mathcal{Q},\sigma'}|  J^\mu(0) |{P,\mathcal{Q},\sigma}\rangle =i\mathcal{Q} \bar{u}_{X\sigma'} \Gamma^{\mu} u_{\rm i\sigma}  \quad \textrm{with}  \\
& & \Gamma^{\mu} = \gamma^{\mu} \mathcal{F}_1\left(q^2\right)+\frac{\sigma^{\mu \nu} q_{\nu}}{2M}\mathcal{F}_2\left(q^2\right) -iq^{\mu} \mathcal{F}_3\left(q^2\right)+\gamma^{\mu} \gamma^5 \mathcal{F}_1^{5}\left(q^2\right)-i\frac{q^{\mu}}{M}\gamma^5  \mathcal{F}_3^{5}\left(q^2\right),
\end{eqnarray}
where we have used the fact that $1$, $\gamma^{\mu}$, $\sigma^{\mu \nu}$, $\gamma^{\mu} \gamma^5$ and $\gamma^5$ form the complete sets of $4\times 4$ $\gamma$ matrices. Among all these form factors, $\mathcal{F}_1\left(q^2\right)$ and $\mathcal{F}_2\left(q^2\right)$ are the Dirac and Pauli form factors, respectively. They are related to the vector current exchange. $\mathcal{F}_1^{5}\left(q^2\right)$ and $\mathcal{F}_3^{5}\left(q^2\right)$ are the axial form factors related to axial vector current. $\mathcal{F}_3\left(q^2\right)$ usually vanishes if the current is conserved. It is easy to see that in our present framework, the $\sigma^{\mu \nu}$ component is missing, and thus the $\mathcal{F}_2$ is zero\footnote{We wish to thank S. Brodsky and G. Teramond for pointing out that a nonzero nucleon Pauli form factor $\mathcal{F}_2$ can be obtained with further assumptions (e.g., see their Erice talk and Ref.~\cite{Brodsky:2008pg}. ). }. In ref.~\cite{Abidin:2009hr}, where non-vanishing Pauli form factor $\mathcal{F}_2$ is obtained,  a new $\sigma^{\mu \nu}$ term has to be introduced into the action.

Before we calculate the form factors from the current expectation value, let us take a look at how current conservation is satisfied. The current conservation condition can be written as
\begin{eqnarray}
& &q_\mu \langle {P_X,\mathcal{Q},\sigma'}|  J^\mu(0) |{P,\cal{Q},\sigma}\rangle \nonumber\\
&\sim&
\int_{0}^{1/\Lambda} d z z^{2}\left[q \textrm{K}_1(q z) J_{\tau-2}(M z)J_{\tau-2}(M z)\bar{u}_{X\sigma'} q\hspace{-6pt}\slash P_{+}u_{\rm i\sigma}\right.\nonumber\\
& & \ \ \ -iq^2 \textrm{K}_0(q z) J_{\tau-2}(M z) J_{\tau-1}(M z)\bar{u}_{X\sigma'} P_{+}u_{\rm i\sigma}\nonumber\\
& & \ \ \ + q  \textrm{K}_1(q z)J_{\tau-1}(M z) J_{\tau-1}(M z)\bar{u}_{X\sigma'} q\hspace{-6pt}\slash P_{-}u_{\rm i\sigma}\nonumber\\
& &\left. \ \ \ +iq^2  K_0(q z)J_{\tau-1}(M z) J_{\tau-2}(M z)\bar{u}_{X\sigma'} P_{-}u_{\rm i\sigma}\right].
\end{eqnarray}
Using the Dirac equation, one can simplify above expression and obtain
\begin{eqnarray}
& &q_\mu \langle {P_X,\mathcal{Q},\sigma'}|  J^\mu(0) |{P,\cal{Q},\sigma}\rangle \nonumber\\
&\sim&
i\int_{0}^{1/\Lambda} d z z^{2}\left[ q M \textrm{K}_1(q z) J_{\tau-2}(M z)J_{\tau-2}(M z)\bar{u}_{X\sigma'} P_{+}u_{\rm i\sigma}\right.\nonumber\\
& &\ \ \ \ -q M \textrm{K}_1(q z) J_{\tau-2}(M z)J_{\tau-2}(M z)\bar{u}_{X\sigma'} P_{-}u_{\rm i\sigma}\nonumber\\
& & \ \ \ \ -q^2 \textrm{K}_0(q z) J_{\tau-2}(M z) J_{\tau-1}(M z)\bar{u}_{X\sigma'} P_{+}u_{\rm i\sigma}\nonumber\\
& & \ \ \ \ +q M \textrm{K}_1(q z)J_{\tau-1}(M z) J_{\tau-1}(M z)\bar{u}_{X\sigma'} P_{-}u_{\rm i\sigma}\nonumber\\
& & \ \ \ \ - q M \textrm{K}_1(q z)J_{\tau-1}(M z) J_{\tau-1}(M z)\bar{u}_{X\sigma'} P_{+}u_{\rm i\sigma}\nonumber\\
& &\left. \ \ \  \ +q^2  \textrm{K}_0(q z)J_{\tau-1}(M z) J_{\tau-2}(M z)\bar{u}_{X\sigma'} P_{-}u_{\rm i\sigma} \right]
\end{eqnarray}
Using the following identities,
\begin{eqnarray}
\frac{d}{dx}\left[x^\nu K_\nu(x)\right]=-x^\nu K_{\nu-1}(x) ,\,   \frac{d}{dx}[x^{\nu} J_\nu(x)]=x^\nu J_{\nu-1}(x),\, \frac{d}{dx}\left[x^{-\nu}J_\nu(x)\right]=-x^{-\nu}J_{\nu+1}(x)
\end{eqnarray}
one can easily show that
\begin{eqnarray}
& &\int_0^{1/\Lambda} z^2K_0(q z)J_{\tau-2}(M z)J_{\tau-1}(M z) dz \nonumber\\
&=&-\frac{1}{q\Lambda^2}K_1(q/\Lambda)J_{\tau-2}(M/\Lambda)J_{\tau-1}(M/\Lambda)\nonumber\\
& &+\frac{1}{q}\int_0^{1/\Lambda} z^2K_1(q z) \left[M J_{\tau-2}(M z)J_{\tau-2}(M z)-M J_{\tau-1}(M z)J_{\tau-1}(M z)\right] dz
\end{eqnarray}
and eventually
\begin{equation}
q_\mu \langle {P_X,\mathcal{Q},\sigma'}|  J^\mu(0) |{P,\cal{Q},\sigma}\rangle\sim \textrm{K}_{1}\left(\frac{q}{\Lambda}\right)J_{\tau-1}\left(\frac{M}{\Lambda}\right)J_{\tau-2}\left(\frac{M}{\Lambda}\right). \label{cc}
\end{equation}
This indicates that the current is conserved when
\begin{equation}
M=\beta_{\tau-2, k} \Lambda \quad \textrm{or} \quad M=\beta_{\tau-1, k} \Lambda, \label{mas}
\end{equation}
where $\beta_{\tau-2, k}$ and $\beta_{\tau-1, k}$ are $k$-th zeros
of $J_{\tau-2}\left(\frac{M}{\Lambda}\right)$ and
$J_{\tau-1}\left(\frac{M}{\Lambda}\right)$, respectively. This is
essentially equivalent to the mass spectrum found in
Ref.~\cite{deTeramond:2005su} by requiring vanishing chiral spinor
wavefunction on the hard wall located at $r_0=\Lambda R^2$.

Furthermore, we would like to comment that in large $q^2$ limit, the current conservation is trivially satisfied when one set  the upper limit of $z$-integral as $\infty$, where we find
\begin{equation}
q_\mu \langle {P_X,\mathcal{Q},\sigma'}|  J^\mu(0) |{P,\cal{Q},\sigma}\rangle\sim i
\left(\bar{u}_{X\sigma'} P_{+}u_{\rm i\sigma} - \bar{u}_{X\sigma'} P_{-}u_{\rm i\sigma}\right)\mathcal{I},
\end{equation}
where $\mathcal{I}$ is found to be
\begin{eqnarray}
\mathcal{I} &=& \frac{2\left(\tau-1\right) }{M}\left(\frac{M^2}{q^2}\right)^{\tau -1}\left[\left. _{2}\textrm{F}_{1}\right. \left(\tau-\frac{3}{2}, \tau; 2\tau-3;\frac{-4M^2}{q^2}\right) -\left. _{2}\textrm{F}_{1}\right. \left(\tau-\frac{1}{2}, \tau; 2\tau-2;\frac{-4M^2}{q^2}\right) \right]\nonumber \\
& & -\frac{2\tau}{M}\left(\frac{M^2}{q^2}\right)^{\tau }\left. _{2}\textrm{F}_{1}\right. \left(\tau-\frac{1}{2}, \tau+1; 2\tau-1;\frac{-4M^2}{q^2}\right).
\end{eqnarray}
Using Taylor expansions of the Hypergeometric functions, one can easily show that $\mathcal{I}=0$.
\begin{figure}[tbp]
\begin{center}
\includegraphics[width=8cm]{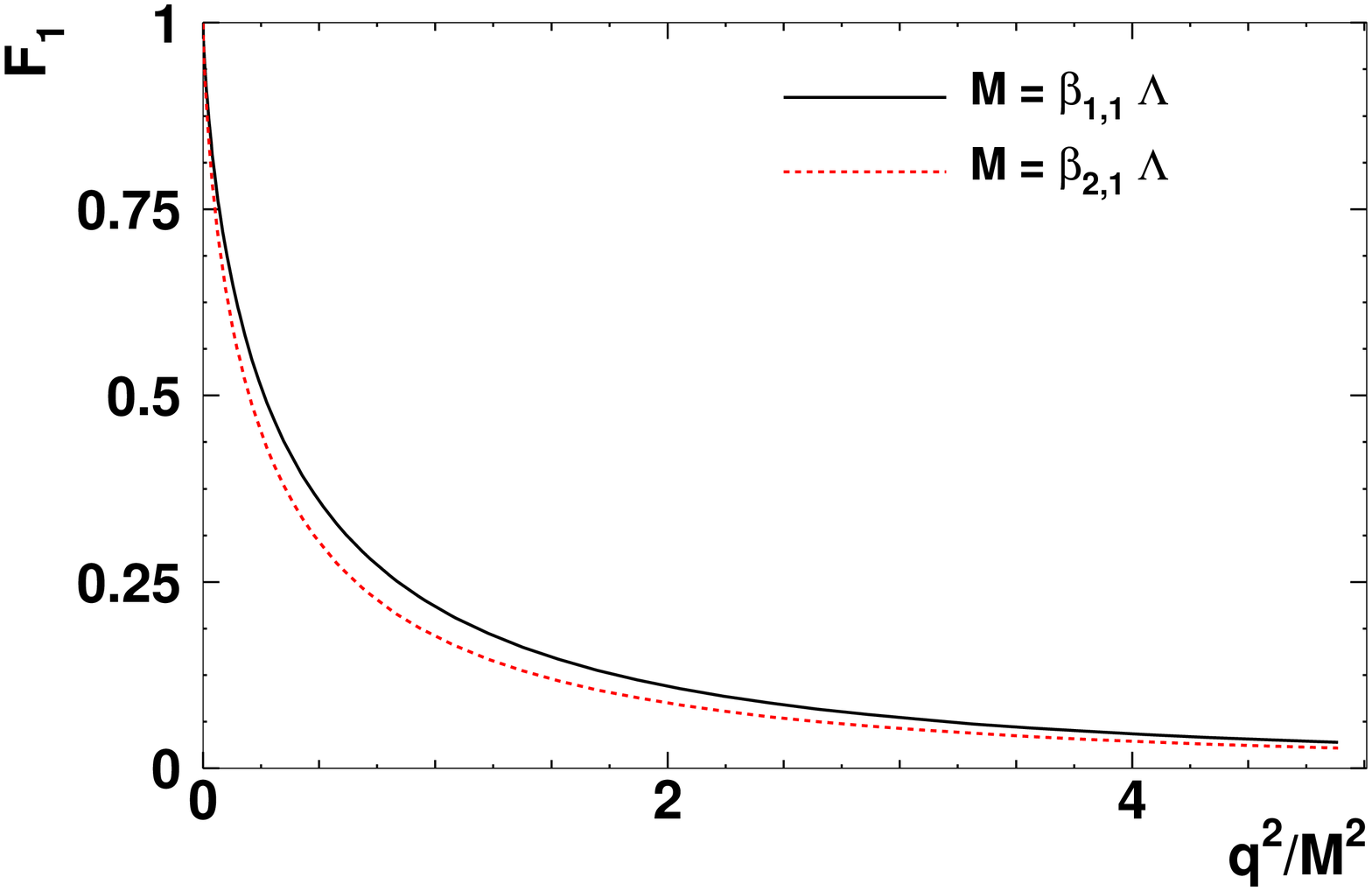}
\includegraphics[width=8cm]{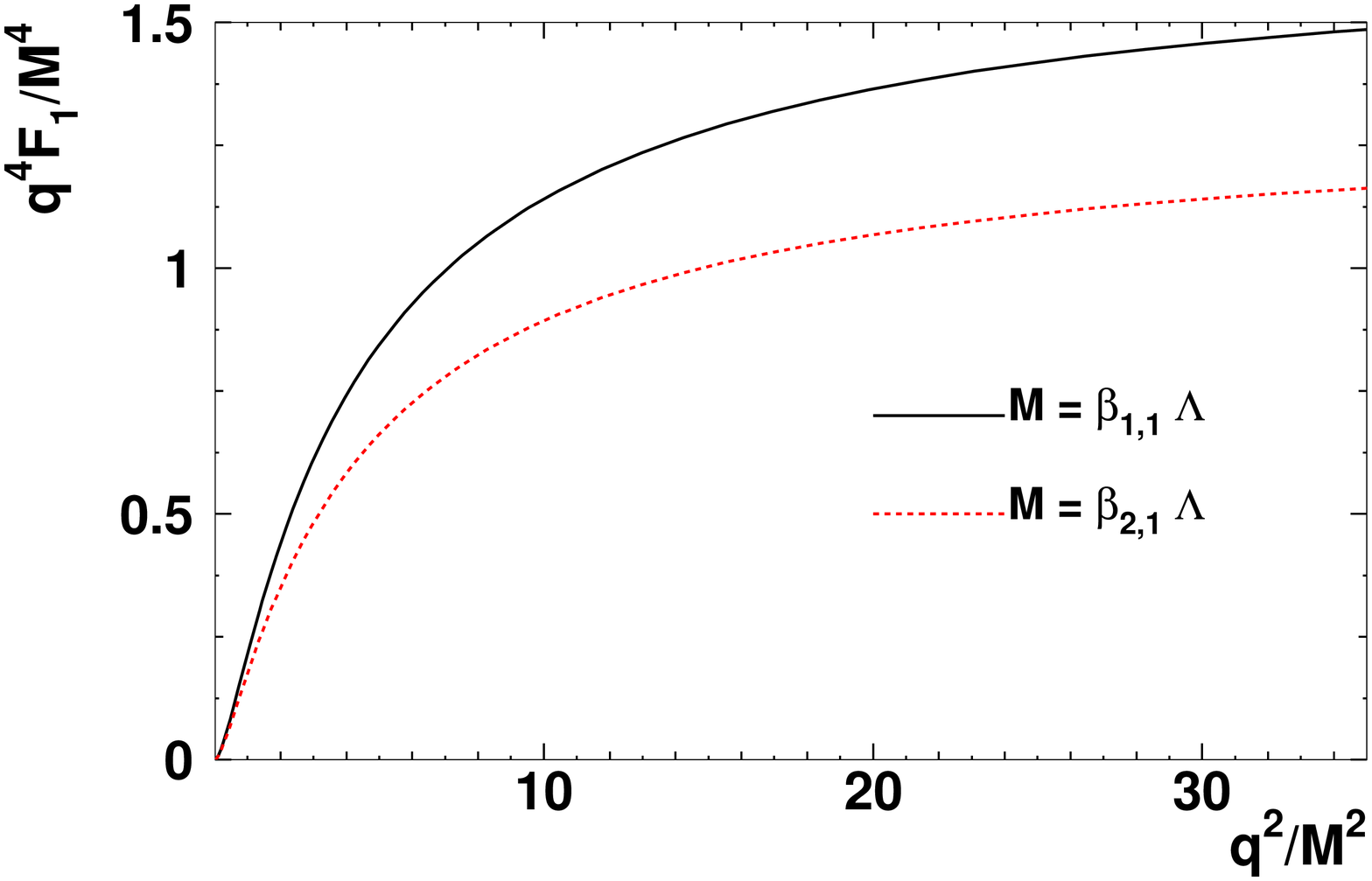}
\includegraphics[width=8cm]{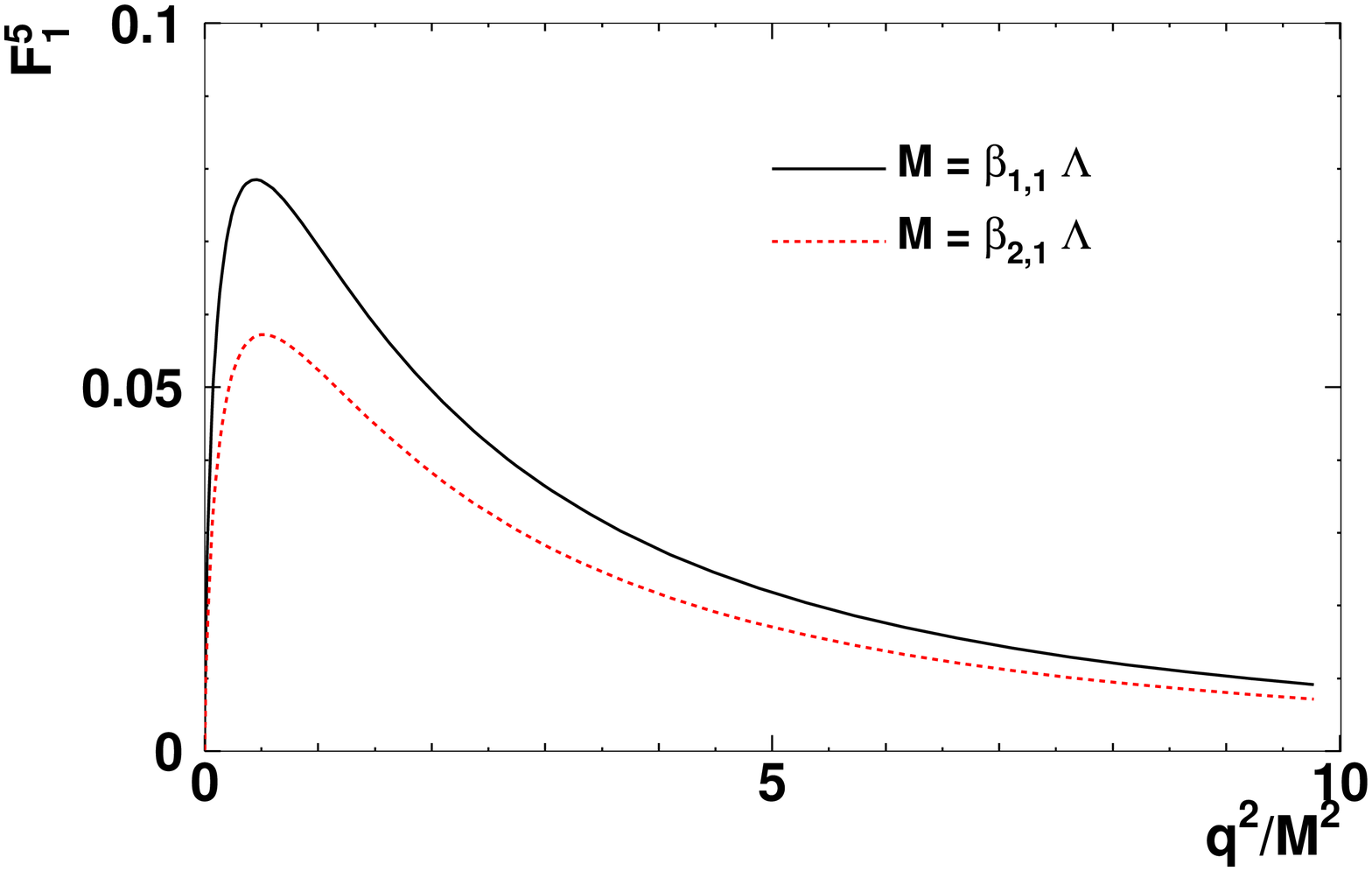}
\includegraphics[width=8cm]{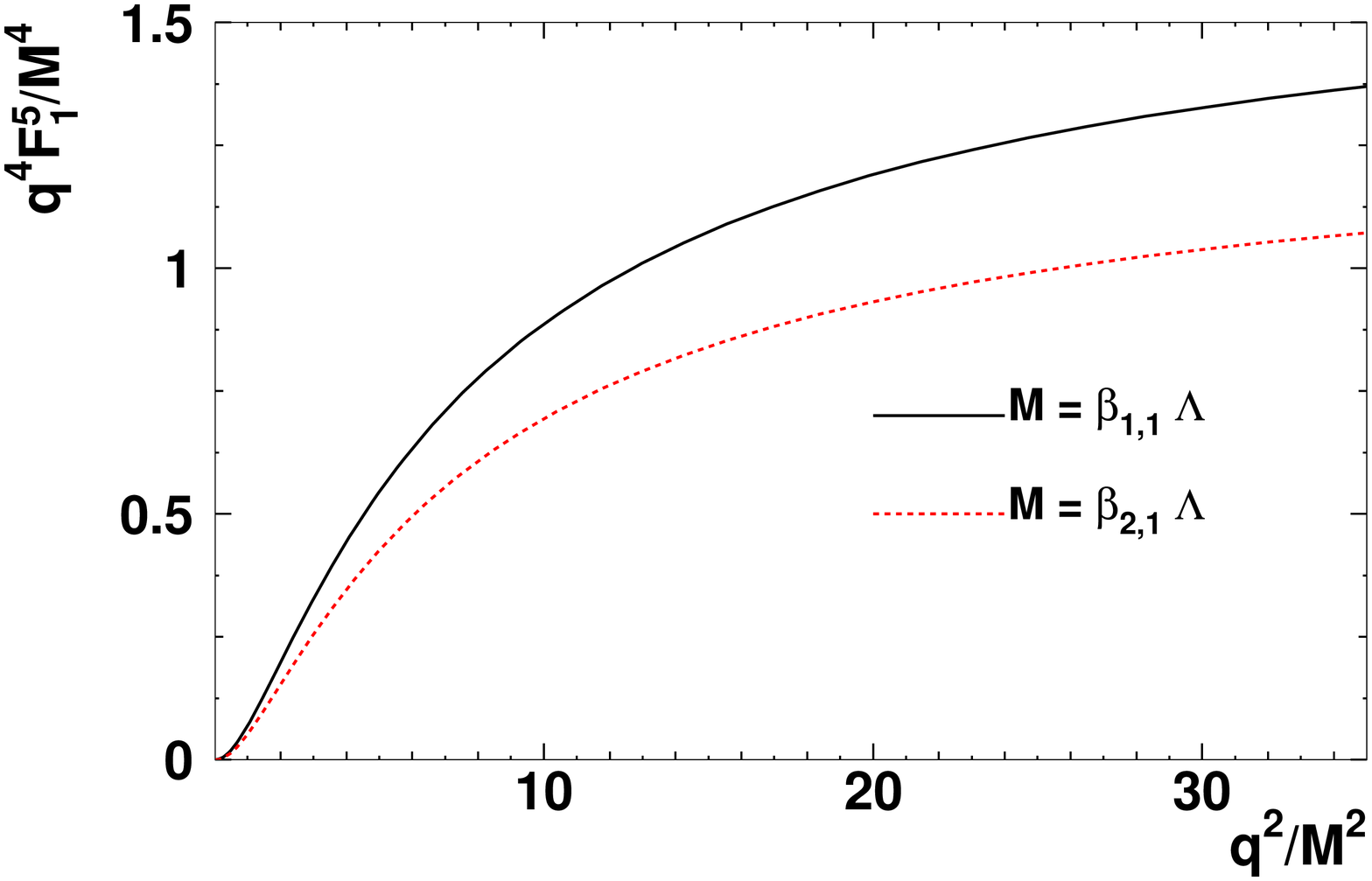}
\includegraphics[width=8cm]{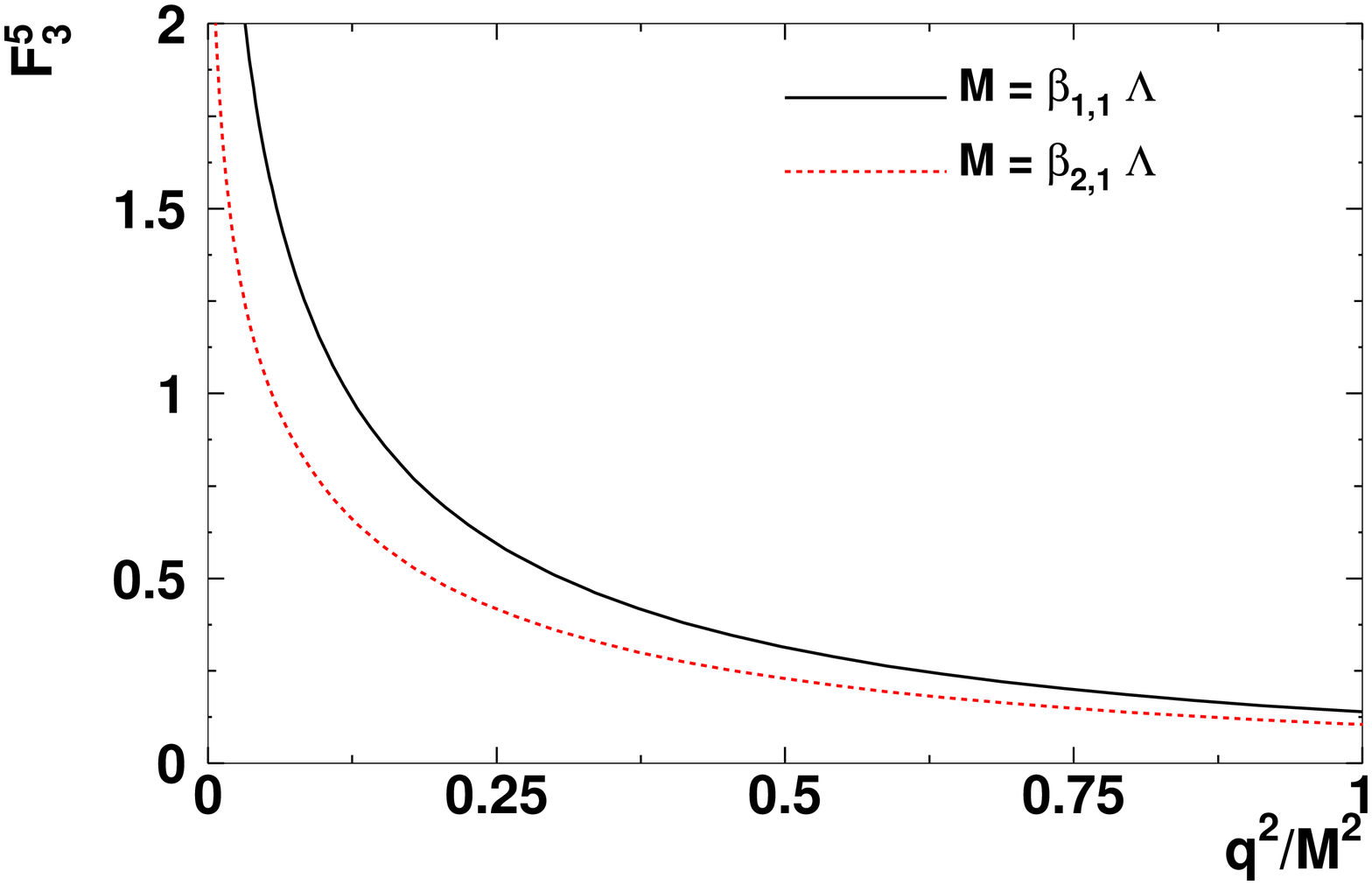}
\includegraphics[width=8cm]{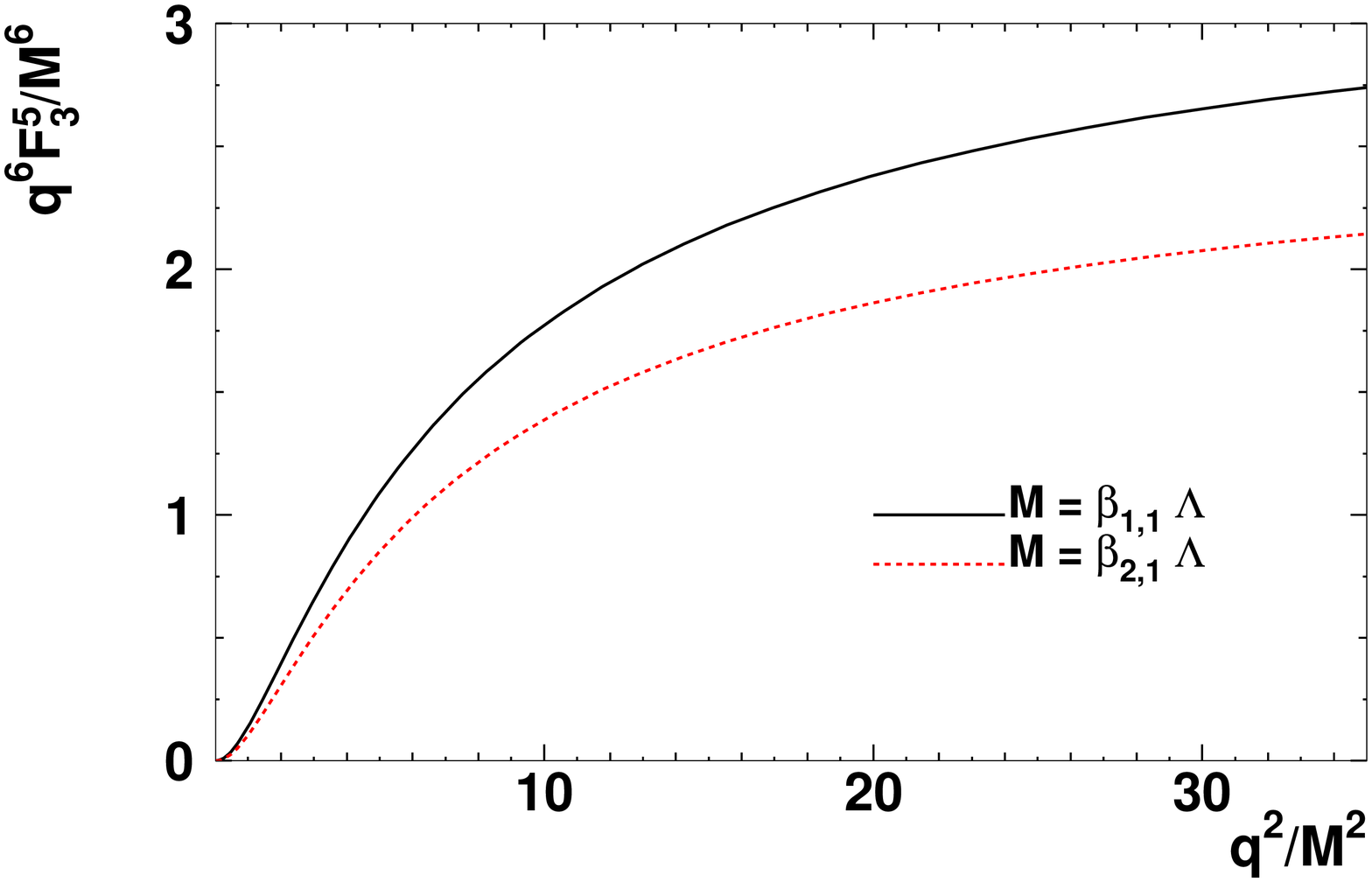}
\end{center}
\caption[*]{Illustration of the $\mathcal{F}_1(q^2)$, $\mathcal{F}^5_1(q^2)$ and $\mathcal{F}^5_3(q^2)$, where we have normalized $\mathcal{F}_1(0)=1$. We also set $\tau=3$ and $M=\beta_{1,1} \Lambda$ or $M=\beta_{2,1}\Lambda$ where $\beta_{1,1}$ and $\beta_{2,1}$ are the first zero root of $J_{1}(\beta)$ and $J_{2}(\beta)$, respectively. }
\label{formfactors}
\end{figure}
\subsection{Elastic form factors in large $q^2$ limit}
Assuming $q\gg \Lambda$, one can set the upper limit of $\textrm{d} z$ integral as $\infty$ and thus obtain
\begin{eqnarray}
\mathcal{F}_1\left(q^2\right) &=&|c^{\prime}|^2\frac{\Lambda}{M}\left(\tau -1\right) \left(\frac{M^2}{q^2}\right)^{\tau -1}\left. _{2}\textrm{F}_{1}\right. \left(\tau-\frac{3}{2}, \tau; 2\tau-3;\frac{-4M^2}{q^2}\right) \nonumber \\
&&+ |c^{\prime}|^2\frac{\Lambda}{M} \tau \left(\frac{M^2}{q^2}\right)^{\tau}  \left. _{2}\textrm{F}_{1}\right. \left(\tau-\frac{1}{2}, \tau+1; 2\tau-1;\frac{-4M^2}{q^2}\right)\\
\mathcal{F}_2\left(q^2\right)&=& 0 \quad \textrm{and} \quad \mathcal{F}_3\left(q^2\right)=0
\end{eqnarray}
and
\begin{eqnarray}
\mathcal{F}_1^5\left(q^2\right) &=&|c^{\prime}|^2\frac{\Lambda}{M}\left(\tau -1\right) \left(\frac{M^2}{q^2}\right)^{\tau -1}\left. _{2}\textrm{F}_{1}\right. \left(\tau-\frac{3}{2}, \tau; 2\tau-3;\frac{-4M^2}{q^2}\right) \nonumber \\
&&- |c^{\prime}|^2\frac{\Lambda}{M} \tau \left(\frac{M^2}{q^2}\right)^{\tau}  \left. _{2}\textrm{F}_{1}\right. \left(\tau-\frac{1}{2}, \tau+1; 2\tau-1;\frac{-4M^2}{q^2}\right)\\
\mathcal{F}_3^5\left(q^2\right) &=& 2|c^{\prime}|^2\frac{\Lambda}{M}\left(\tau -1\right) \left(\frac{M^2}{q^2}\right)^{\tau }\left. _{2}\textrm{F}_{1}\right. \left(\tau-\frac{1}{2}, \tau; 2\tau-2;\frac{-4M^2}{q^2}\right).
\end{eqnarray}
At large $q^2$ limit, we find that $\mathcal{F}_1\left(q^2\right)\simeq\mathcal{F}_1^5\left(q^2\right)\simeq |c^{\prime}|^2\frac{\Lambda}{M}\left(\tau -1\right) \left(\frac{M^2}{q^2}\right)^{\tau -1}$ and  $\mathcal{F}_3^5\left(q^2\right)\simeq 2|c^{\prime}|^2\frac{\Lambda}{M}\left(\tau -1\right) \left(\frac{M^2}{q^2}\right)^{\tau }$.

\subsection{Elastic form factors in small $q^2$ limit}
In small $q^2$ limit, we expand the Bessel functions $\textrm{K}_{0,1}(q^2)$ up to $q^2 \log q^2$ but neglect $q^2$ terms. It is then straightforward to evaluate the $\textrm{d}z$ integral which yields
\begin{eqnarray}
\mathcal{F}_1\left(q^2\right) &=&|c^{\prime}|^2\frac{M}{2\Lambda}\left[J_{\tau -2}\left(\frac{M}{\Lambda}\right)J_{\tau -1}^{\prime}\left(\frac{M}{\Lambda}\right)-J_{\tau -1}\left(\frac{M}{\Lambda}\right)J_{\tau -2}^{\prime}\left(\frac{M}{\Lambda}\right)\right]\nonumber\\
&& +2|c^{\prime}|^2\left(\frac{M}{2\Lambda}\right)^{2\tau-1}\frac{q^2}{M^2}\ln\left(\frac{q}{\Lambda}\right)
\frac{\ _2F_3(\tau-\frac{3}{2},\tau;\tau-1,\tau+1,2\tau-3;-M^2/\Lambda^2)}{2\tau\Gamma(\tau-1)^2}\nonumber\\
&&+\frac{1}{2}|c^{\prime}|^2
\left(\frac{M}{2\Lambda}\right)^{2\tau+1}\frac{q^2}{M^2}\ln\left(\frac{q}{\Lambda}\right)
\frac{\
_2F_3(\tau-\frac{1}{2},\tau+1;\tau,\tau+2,2\tau-1;-M^2/\Lambda^2)}{2(\tau+1)\Gamma(\tau)^2} \\
\mathcal{F}_2\left(q^2\right)&=& 0 \quad \textrm{and} \quad \mathcal{F}_3\left(q^2\right)=0
\end{eqnarray}
and
\begin{eqnarray}
\mathcal{F}_1^5\left(q^2\right) &=&\frac{1}{2}|c^{\prime}|^2 J_{\tau -2}\left(\frac{M}{\Lambda}\right) J_{\tau -1}\left(\frac{M}{\Lambda}\right)\nonumber \\
&&+2|c^{\prime}|^2
\left(\frac{M}{2\Lambda}\right)^{2\tau-1}\frac{q^2}{M^2}\ln(q/\Lambda)
\frac{\ _2F_3(\tau-\frac{3}{2},\tau;\tau-1,\tau+1,2\tau-3;-M^2/\Lambda^2)}{2\tau\Gamma(\tau-1)^2}\nonumber\\
& &-\frac{1}{2}|c^{\prime}|^2 \left(\frac{M}{2\Lambda}\right)^{2\tau+1}\frac{q^2}{M^2}\ln(q/\Lambda)
\frac{\
_2F_3(\tau-\frac{1}{2},\tau+1;\tau,\tau+2,2\tau-1;-M^2/\Lambda^2)}{2(\tau+1)\Gamma(\tau)^2}
\end{eqnarray}
together with
\begin{eqnarray}
{\mathcal{F}}_3^5(q^2)&=&-|c^{\prime}|^2
\left(\frac{M}{\Lambda}\right)^{2\tau-1}2^{2-2\tau}\ln(q/\Lambda)
\frac{\
_1F_2(\tau-\frac{1}{2};\tau+1,2\tau-2;-M^2/\Lambda^2)}{\Gamma(\tau-1)\Gamma(\tau+1)}
\end{eqnarray}

\begin{figure}[tbp]
\begin{center}
\includegraphics[width=8cm]{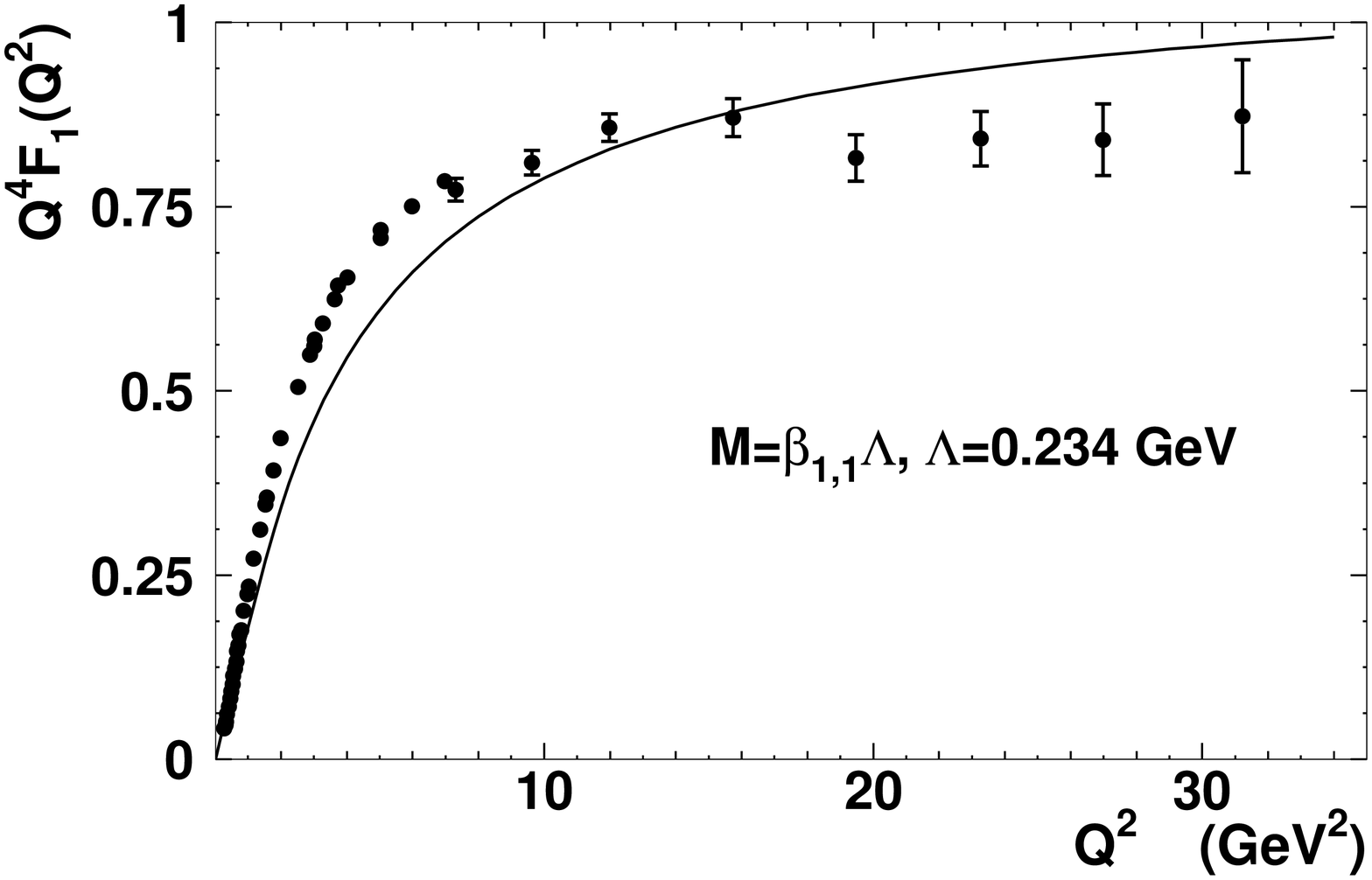}
\end{center}
\caption[*]{Predictions for $Q^4F_1^p(Q^2)$ in the hard wall
model. The data points are taken from \cite{Diehl:2005wq}}
\label{formfit}
\end{figure}

In the end, one can use numerical methods and evaluate all these
form factors with chosen $\tau$ and ratio $M/\Lambda$, then plot
them in terms of functions of $q^2/M^2$(See
Figure.~(\ref{formfactors})). According to the power counting
rule, we set $\tau =3$ for now. It is easy to see that above form
factors give rise to logarithmic divergent charge radii for the
charged hadron. This is peculiar in the hard wall model and will
be cured in our following-up phenomenological studies\cite{phe}.
Besides, we also compare our results of $Q^4F_1^p(Q^2)$ with
experimental data, which are shown in Figure.~(\ref{formfit}).

\section{Conclusion} \label{conc}
Using gauge/string duality, we have calculated the structure functions as well as the form factors of
a spin-$\frac{1}{2}$ hadron. Especially the polarized structure functions and parity violating
structure functions are new. We find that the Burkhardt-Cottingham sum rule is also true in our
present calculation when the small-$x$ contribution to $g_2$ is negligible. However, the situation
for the $g_1$ structure function is more subtle and complicated. We conjecture that there should be an axial regge contribution to $g_1$ at small-$x$ which may indicate that most of the hadron spin is carried by small-$x$ partons. The phenomenological application of above calculation is very appealing and will be available soon\cite{phe}.

\begin{acknowledgments}
We acknowledge inspiring discussions with S. Brodsky, V. Koch, Y.
Kovchegov, A. Mueller, J. Qiu, G. Teramond and F. Yuan. We would
like to thank Y. Hatta for enormous communication and stimulating
discussions during the preparation of this paper. J.G.
acknowledges financial support by the National Natural Science
Foundation of China under Grant No.10525523. B.X. is supported by
the Director, Office of Energy Research, Office of High Energy and
Nuclear Physics, Divisions of Nuclear Physics, of the U.S.
Department of Energy under Contract No. DE-AC02-05CH11231.

\end{acknowledgments}

\end{document}